# Carefully choose the baseline: Lessons learned from applying XAI attribution methods for regression tasks in geoscience

by

Antonios Mamalakis[1*], Elizabeth A. Barnes[1], Imme Ebert-Uphoff [2,3]

[1] Department of Atmospheric Science, Colorado State University, Fort Collins, CO
[2] Department of Electrical and Computer Engineering, Colorado State University, Fort Collins, CO
[3] Cooperative Institute for Research in the Atmosphere, Colorado State University, Fort Collins, CO

*email: amamalak@colostate.edu

**Abstract**
Methods of eXplainable Artificial Intelligence (XAI) are used in geoscientific applications to gain insights into the decision-making strategy of Neural Networks (NNs) highlighting which features in the input contribute the most to a NN prediction. Here, we discuss our "lesson learned" that the task of attributing a prediction to the input does not have a single solution. Instead, the attribution results and their interpretation depend greatly on the considered baseline (sometimes referred to as "reference point") that the XAI method utilizes; a fact that has been overlooked so far in the literature. This baseline can be chosen by the user or it is set by construction in the method's algorithm – often without the user being aware of that choice. We highlight that different baselines can lead to different insights for different science questions and, thus, should be chosen accordingly. To illustrate the impact of the baseline, we use a large ensemble of historical and future climate simulations forced with the SSP3-7.0 scenario and train a fully connected NN to predict the ensemble- and global-mean temperature (i.e., the forced global warming signal) given an annual temperature map from an individual ensemble member. We then use various XAI methods and different baselines to attribute the network predictions to the input. We show that attributions differ substantially when considering different baselines, as they correspond to answering different science questions. We conclude by discussing some important implications and considerations about the use of baselines in XAI research.

**Significance Statement**
In recent years, methods of eXplainable Artificial Intelligence (XAI) have found great application in geoscientific applications, as they can be used to attribute the predictions of Neural Networks (NNs) to the input and interpret them physically. Here, we highlight that the attributions – and the physical interpretation – depend greatly on the choice of the baseline; a fact that has been overlooked in the literature. We illustrate this dependence for a specific climate task, where a NN is trained to predict the ensemble- and global-mean temperature (i.e., the forced global warming signal) given an annual temperature map from an individual ensemble member. We show that attributions differ substantially when considering different baselines, as they correspond to answering different science questions. We conclude by discussing some important implications and considerations about the use of baselines in XAI research.

## 1. Introduction

EXplainable Artificial Intelligence (XAI) aims to provide insights about the decision-making process of AI models and has been increasingly applied to the geosciences (e.g., Toms et al., 2021; Ebert-Uphoff and Hilburn, 2020; Hilburn et al., 2021; Barnes et al., 2019, 2020; Mayer and Barnes, 2021; Keys et al., 2021; Sonnewald and Lguensat, 2021). XAI methods show promising results in calibrating model trust, and assisting in learning new science (see for example, McGovern et al., 2019; Toms et al., 2020; Mamalakis et al., 2022a). A popular subcategory of XAI is the so-called local attribution methods, which compute the attribution of a model's prediction to the features in the input. The final product typically comes in the form of a heatmap in the shape of the original input. Because of the complex architecture of state-of-the-art AI models (e.g., neural networks; NNs), the "attribution task" can be challenging, and many XAI methods have been shown to not honor desired properties (Kindermans et al., 2017; Ancona et al., 2018; 2019; Rudin, 2019; Dombrowski et al., 2020). Inter-comparison studies have shown that the faithfulness and the comprehensibility of the attributions depends on the prediction setting and the model architecture, and that no universally optimal method likely exists (Mamalakis et al., 2022b,c).

Apart from the issue of the fidelity of the methods, another very important aspect of the attribution task is that, in regression problems, one needs to define the *baseline* that the attribution is calculated for, i.e., the baseline is the reference point, which the prediction is compared to so it can be interpreted. Without having defined a baseline, the attribution task cannot be solved - it is an ill-defined task. Defining a baseline is necessary not only when explaining a complex AI model, but also for explaining very simple models: thus, defining a baseline is tied to the attribution task itself and not the model. To illustrate the above, let us consider the simple case of a linear model with no constant term: $y = f(\mathbf{x}) = \sum_i w_i x_i$, and let us suppose that we want to explain/attribute a prediction $y_n$ to the input features. A trivial solution to the attribution task for this case seems to be that any input feature $x_i$ contributes $w_i x_{i,n}$ to the prediction $y_n$. However, this is a correct attribution rule only if we assume a baseline $\hat{\mathbf{x}} = \mathbf{0}$. Thus, even in cases when the attribution task seems to have exactly one, trivial solution, this solution corresponds to a baseline that has been assumed implicitly. The complete way to attribute the prediction $y_n$ to the input is to consider the *set of solutions*: $w_i(x_{i,n} - \hat{x}_i)$. It can be observed that the attribution depends on the considered baseline and that the previous, trivial solution is simply a special case of the complete set of solutions. Another special case that is typically very informative to scientists is to choose a baseline of the form $\hat{\mathbf{x}} = \mathrm{E}[\mathbf{x}]$. In this case, the attribution would highlight those features in the input for which the deviation from the average state is important for the prediction.

In this paper, our aim is to highlight our lesson learned that the attribution task is highly dependent on the choice of baseline – a fact that seems to have been overlooked so far in applications of XAI to the geosciences. Also, we show by example how the use of different baselines offers an opportunity to answer different science questions. To do so, we use a large ensemble of historical and future climate simulations forced with the SSP3-7.0 climate change scenario (ensemble of 80 members from the climate model CESM2; Rodgers et al., 2021) and train a fully connected NN to predict the ensemble- and global-mean temperature (i.e., the forced component of the global mean temperature) given an annual temperature map from an individual ensemble member. We then use XAI methods and different baselines to attribute the network predictions to the input. We show that attributions differ substantially when considering different baselines, as they correspond to answering different science questions.

In section 2, we provide details about the data, prediction task and methods, and in section 3 we present our results. Section 4 discusses some considerations about the use of baselines in XAI research and in section 5 we provide a summary of the key points of our study.

## 2. Data and methodology
### 2.1. Data

We use yearly-mean surface air temperature from the CESM2-Large Ensemble Community project (Rodgers et al., 2021), spanning the years from 1850 to 2100 (publicly available at https://www.cesm.ucar.edu/projects/community-projects/LENS2/data-sets.html). Historical forcing is applied to the climate system over the 1850-2014 period and the SSP3-7.0 climate change scenario is applied over the 2015-2100 period. We use the output of 80 members bi-linearly re-gridded to a 2.5° by 2.5° resolution from an approximate 1° by 1° resolution to reduce dimensionality of the prediction task.

### 2.2. Prediction task and network architecture

Given a yearly-mean temperature map as an input (10,512 pixels), we train a fully connected NN to predict the ensemble- and global-mean temperature of the same year as the input map (see graphic in Figure 1). This task requires the NN to recognize the forced signal in the temperature field (i.e., the signal originating from natural forcings, such as solar radiation and volcanic eruptions, or anthropogenic forcings, such as changes in greenhouse gases concentrations, tropospheric aerosols, land use, etc.) in order to estimate the forced global mean temperature, while ignoring any internal variability signals that may be present in the input (e.g., an active El Niño-Southern Oscillation; ENSO). Thus, the NN needs to be able to separate the forced temperature response from that of the internal climate "noise".

Our network consists of two hidden layers with eight and five neurons each (with ReLU activation functions) and one output neuron with no activation. Dropout (with probability of 0.3) and ridge regularization (with a regularization factor of 0.005) are applied in the input layer to avoid overfitting, and training is performed using the Adam optimizer (Kingma and Ba, 2017), a batch size of 32 and a learning rate of 0.0001. The mean squared error is used as the loss function during training. We train on 70 members and test on the remaining 10 members. The performance of the network is satisfying with a mean absolute error on the order of 0.068° C for the testing data (and an $R^2$ on the order of 99%).

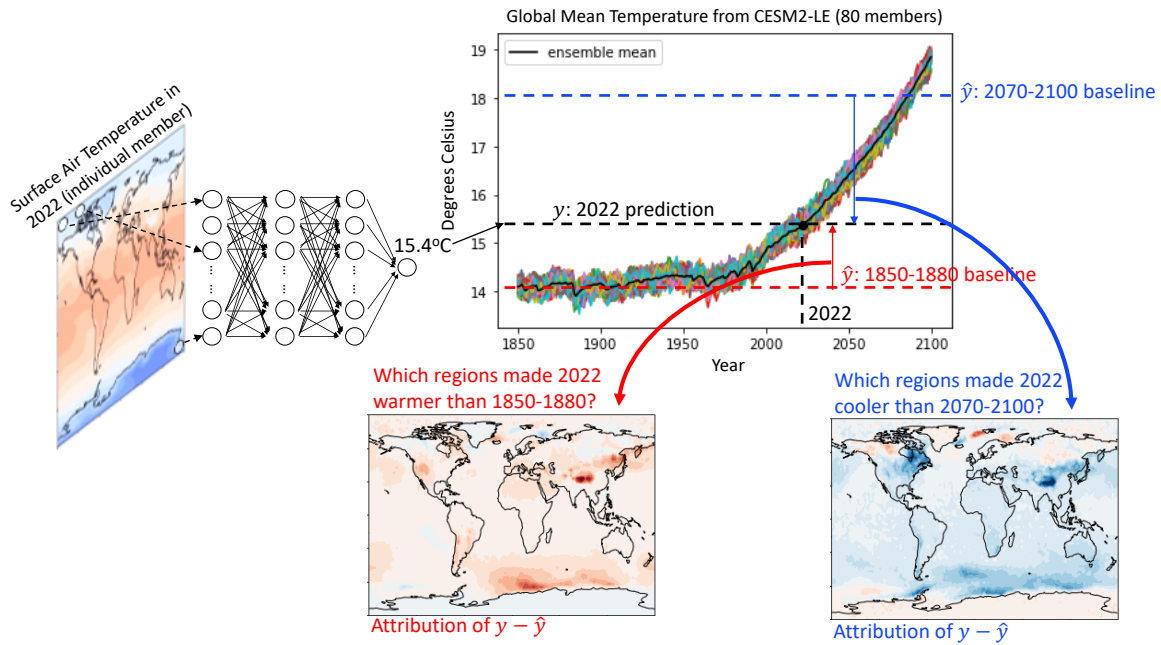

**Figure 1.** Schematic representation of the prediction task of the study and of the use of baselines to gain insights into different science questions.

### 2.3. XAI methods

The main goal of the current work is the explanation rather than the prediction itself. To explain the predictions of the network we use two local attribution methods, namely the Integrated Gradients (Sundararajan et al., 2017) and Deep SHAP (Lundberg and Lee, 2017). We have chosen these methods for two main reasons: i) Both methods allow the user to define the baseline for which the attribution is derived, which is the focus of this work and allows us to gain insights into different science questions (see Figure 1). ii) Both methods satisfy the *completeness* property (also referred to as *local accuracy* in Lundberg and Lee, 2017, or *sensitivity-N* in Ancona et al., 2018), which holds that the feature attributions must add up to the difference between the NN output at the current sample $\mathbf{x}_n$ and the baseline $\hat{\mathbf{x}}$. We briefly describe the two methods below.

**Integrated Gradients:** This method (Sundararajan et al., 2017) is a local attribution method that builds on the Input*Gradient method (Shrikumar et al., 2016; 2017). Namely, it aims to account for the fact that in nonlinear problems the gradient is not constant, and thus, the product of the local gradient with the input might not be a good approximation of the input's contribution. The Integrated Gradients method considers a baseline vector $\hat{\mathbf{x}}$, and the attribution for feature $i$ and sample $n$ is equal to the product of the distance of the input from the baseline with the average of the gradients at points along the straight-line path from the baseline to the input:

$$R_{i,n} = (x_{i,n} - \hat{x}_i) * \frac{1}{m} \sum_{j=1}^{m} \frac{\partial \hat{F}}{\partial X_i}\bigg|_{X_i = \hat{x}_i + \frac{j}{m}(x_{i,n} - \hat{x}_i)} \quad (1)$$

where $\hat{F}$ represents the network and $m$ is the number of steps in the Riemann approximation.

**Deep SHAP:** Deep SHAP is a local attribution method that is based on the use of Shapley values (Shapley, 1953) and is specifically designed for deep neural networks (Lundberg and Lee, 2017). The Shapley values originate from the field of cooperative game theory and represent the average expected marginal contribution of each player in a cooperative game, after all possible combinations of players have been considered (Shapley, 1953). Regarding the importance of Shapley values for XAI, it can be shown (Lundberg and Lee, 2017) that across all *additive feature attribution methods* (a general class of local attribution methods that unifies many popular XAI methods), the only method that satisfies all desired properties of local accuracy, missingness and consistency (see Lundberg and Lee, 2017, for details on these properties) emerges when the feature attributions $\varphi_i$ are equal to the Shapley values:

$$\varphi_i = \sum_{S \subseteq M \setminus \{i\}} \frac{|S|!\,(|M| - |S| - 1)!}{|M|} \left[ f_{S \cup \{i\}}(x_{S \cup \{i\}}) - f_S(x_S) \right] \quad (2)$$

where $M$ is the set of all input features, $M\setminus\{i\}$ is the set $M$, but with the feature $x_i$ being withheld, $|M|$ represents the number of features in $M$, and the expression $f_{S \cup \{i\}}(x_{S \cup \{i\}}) - f_S(x_S)$ represents the net contribution (effect) of the feature $x_i$ to the outcome of the model $f$, which is calculated as the difference between the model outcome when the feature $x_i$ is present and when it is withheld. Thus, the Shapley value $\varphi_i$ is the (weighted) average contribution of the feature $x_i$ across all possible subsets $S \subseteq M\setminus\{i\}$. Due to computational constraints, Deep SHAP approximates the Shapley values for the entire network by computing the Shapley values for smaller components of the network and propagating them backwards until the input layer is reached (i.e., implementing Eq. (2) recursively). Deep SHAP may consider different baselines (defined by the user) to replace features $x_i$ with, when they need to be withheld.

### 3. Results

In this section, we focus on attributing a specific prediction of the NN using different baselines. We choose the 2022 temperature map from the 80[th] ensemble member as the input of interest. The attributions based on the Integrated Gradients and the Deep SHAP are shown in Figure 2 for four different baselines. Figure 3 shows the difference between the considered input to the NN and the four baselines. We note that for the considered input, the NN estimates the forced global mean temperature to be 15.4°C, which is almost identical to the true ensemble- and global-mean temperature for 2022.

We first consider the baseline of zero input (see Figures 2a,e and 3a); a very common choice in XAI applications. Reasonably, the NN output that corresponds to a zero input (although such an input was not present in the training or the testing set) is almost 0°C, and thus the question that we aim to answer here is "*which features in the 2022 map made the network predict the temperature to be 15.4°C as opposed to 0°C?*". According to both XAI methods (the two methods provide almost identical results for all the baselines), the features that determined this difference mainly occur over the zone of 55°S-55°N and partially over Antarctica. Thus, this attribution indicates that the NN has learned the basic concept that the temperature of the globe cannot be 0°C because of the heat stored in the tropics and subtropics originating from the solar radiation and the greenhouse effect. Also, note that the majority of the attributions are positive, since all attributions must add up to a positive value of 15.4°C; i.e., recall that the attributions need to add up to the difference of the NN output at the current 2022 map (15.4°C) minus the NN output at the baseline (0°C).

As a second baseline, we consider the temperature map averaged over the 1850-1880 period from the 80$^{th}$ ensemble member. The NN output that corresponds to this baseline is 14.1°C, and thus the question that we aim to answer here is "*which features in the 2022 map made the network predict the temperature to be 15.4°C as opposed to 14.1°C?*" or alternatively "*which regions made 2022 warmer than the period 1850-1880 by 1.3°C?*". The XAI methods highlight positive attributions mostly over land and oceanic regions in the midlatitudes and specifically over the Himalayas, eastern Asia, North and South America, the Southern Ocean and the northern Pacific Ocean. The high latitudes show negligible contribution, despite the high degree of warming that occurs locally (see Figure 3b). We hypothesize that this is due to the high internal variability that is associated with the warming of these regions, and thus, they may not constitute robust predictors for the NN to determine the forced global warming.

# Attribution of the 2022 Network Prediction for Different Baselines

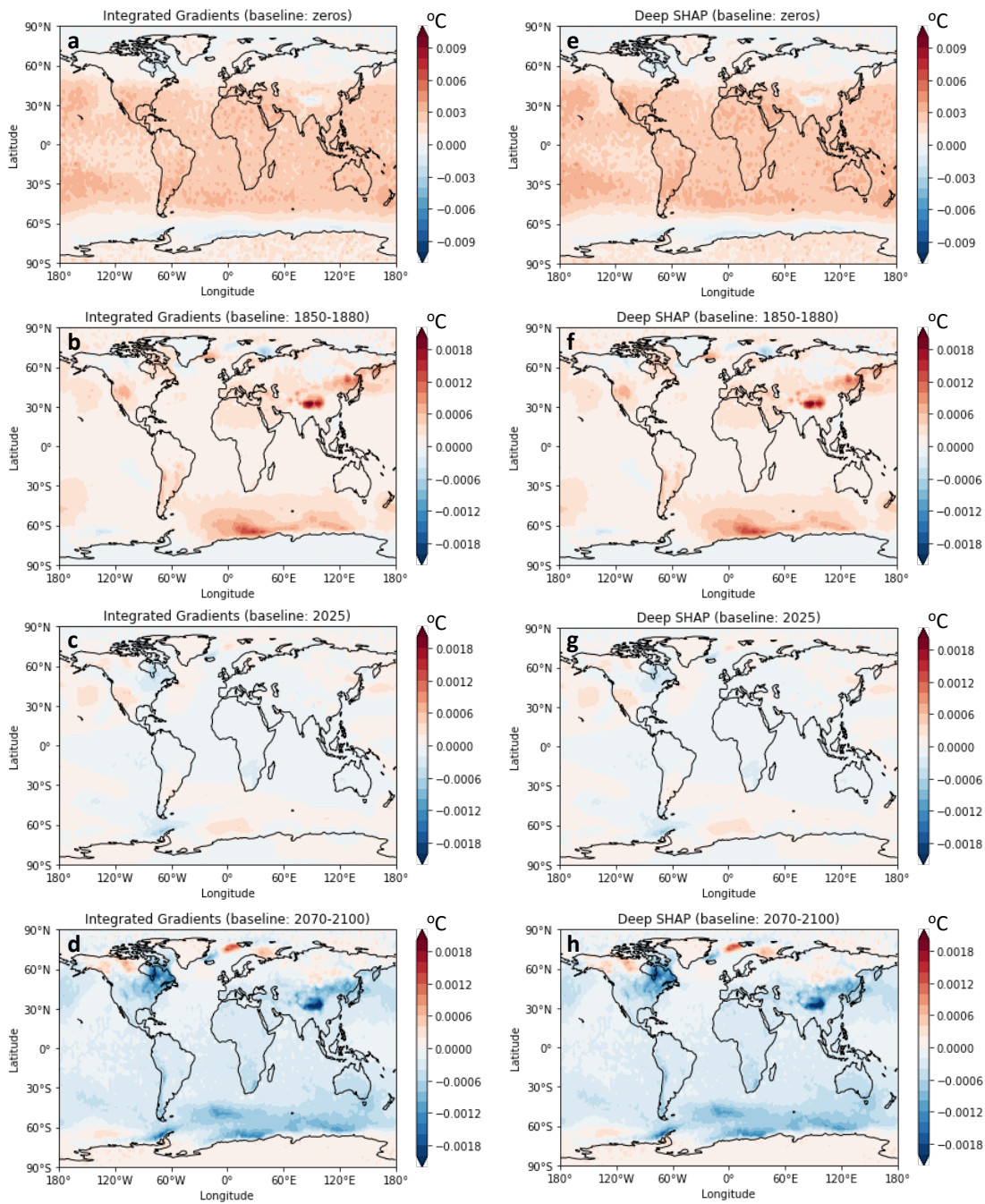

**Figure 2.** Attribution heatmaps (in °C) derived by the methods Integrated Gradients (panels a-d) and Deep SHAP (panels e-h) for the NN prediction for the year 2022 (80[th] member), using four different baselines (different rows). The considered baselines are the zero input (first row), the average temperature over 1850-1880 (from the 80[th] member; 2[nd] row), the temperature in 2025 (from the 80[th] member; 3[rd] row) and the average temperature over 2070-2100 (from the 80[th] member; 4[th] row).

As a third baseline, we next consider the year 2025 (from the 80[th] ensemble member). The NN output for the baseline is 15.7°C, thus, the attributions need to add up to only 15.4°C-15.7°C= –0.3°C. Consequently, attributions are of a lower magnitude in this case. The question answered here is "*which features in the 2022 map made the network predict the temperature to be 15.4°C as* *opposed to 15.7°C?"* or alternatively *"which regions made 2022 cooler than 2025 by 0.3°C?"*. The regions of north Pacific and Southern Oceans are highlighted the most. We note that the strong pattern of an active ENSO over the eastern tropical Pacific in 2025 (see Figure 3c) is not highlighted by the XAI methods. This suggests that the network has learned that ENSO variability constitutes internal variability, thus, not an appropriate predictor of the forced global mean temperature.

Lastly, we consider the average temperature over the period 2070-2100 (from the 80$^{th}$ ensemble member) as the baseline. In this case, we are interested in gaining insights about the regions that made 2022 cooler than the end of the century. Both XAI methods highlight the majority of the globe with negative attribution, since the attributions need to add up to 15.4°C-18°C= –2.6°C. The most important contributors are shown to be the Himalayas, eastern Asia, North America, the Southern Ocean and the northern Pacific Ocean. Similar to the remarks in the case of the 1850-1880 baseline, the high latitudes are shown to contribute only slightly.

The above results make clear that the attribution task depends substantially on the considered baseline. This is true both in terms of the magnitude of the attributions, because of the completeness property (i.e., note the 5-fold difference in the color scale between the top panels and the rest of the panels in Figure 2), but also in terms of spatial patterns. For example, although the Himalayas is shown to be a very important region for distinguishing the forced global warming between the year 2022 and the e.g., 1850-1880 period (see Figure 2b,f), it is not a strong determining factor in the case of a zero baseline (see Figure 2a,e; the same may apply for other regions, e.g., the Southern Ocean). Vice versa, many regions in the deep tropics that are highlighted in the attribution when considering a zero baseline are not highlighted for the other three baselines. This indicates the importance of explicitly declaring the baseline, to avoid misidentifying or overlooking predictive (or causal) factors. At the same time, the above results illustrate that the ill-defined nature of the attribution task is beneficial in that different baselines can be considered to gain insights into different science questions of varying complexity.

# Deviation of Temperature in 2022 from Different Baselines

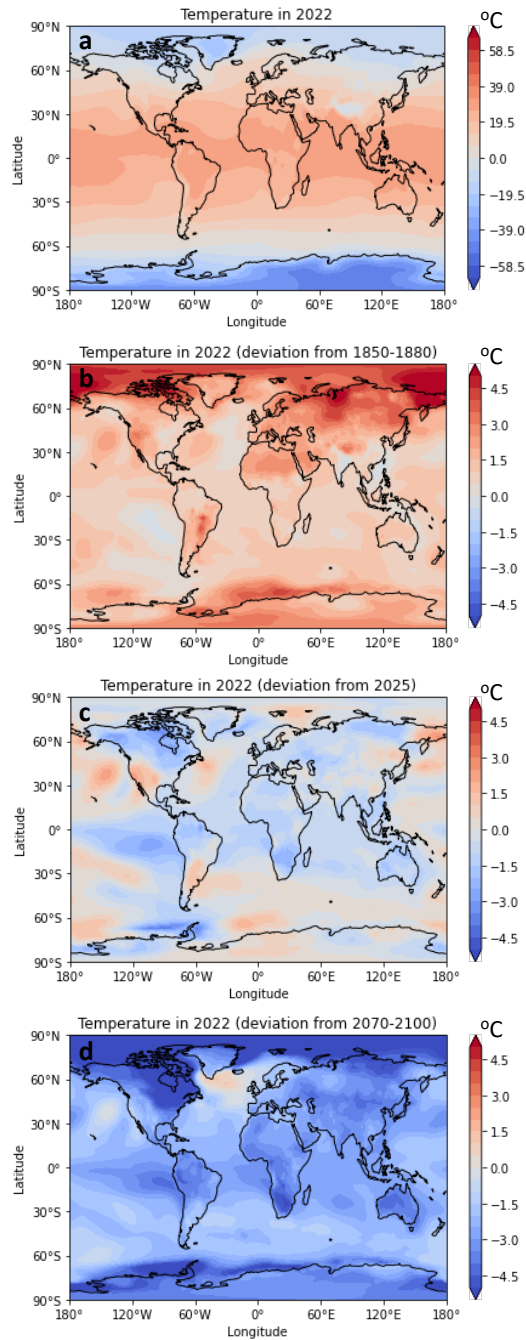

**Figure 3.** Deviation (in °C) of the 2022 surface air temperature from the four baselines used in Figure 2.

## 4. Implications for the use of XAI methods in geoscientific research

The dependency of the attribution task on the baseline highlights a few important considerations for the use of XAI methods in geoscience. First, it means that when comparing explanations using a variety of attribution methods, one must ensure that the same baseline is used for every method to avoid introducing artificial inter-method discrepancies that might be misinterpreted as inter-method variability. For example, in Figure 2, we show that the considered methods Integrated Gradients and Deep SHAP provide almost identical results for the same baseline (likely because of the simple architecture of the NN and the semi-linear nature of the prediction task). However, it would be incorrect to compare their results for different baselines (e.g. comparing panel a with panel f), as they refer to different questions. We highlight this since in current geoscientific research, the baseline is typically not discussed at all when XAI attribution methods are applied.

Second, one needs to keep in mind that although many XAI methods allow the user to choose the baseline (e.g., as the ones used herein), some XAI methods assume specific types of baselines by construction, thus they should be used with caution. For example, unless extra modifications are implemented (Letzgus et al., 2021), the methods Layerwise Relevance Propagation (Bach et al., 2015; Samek et al., 2016) and Input*Gradient (Shrikumar et al.., 2016; 2017) provide attributions using a zero baseline. This implies that a zero-input value is automatically assigned a zero attribution, although the presence of a zero value might be important for the prediction when viewed from a different baseline that might be of interest (this was discussed as the "ignorant to zero input" issue in Mamalakis et al., 2022c).

Lastly, we note that XAI baselines are also very relevant and impactful in classification settings, but they should be used differently than in regression settings. We showed here that in regression settings, different baselines form the necessary decision boundaries (i.e., the questions "…, as opposed to…") for the user to understand why certain decisions were made. In classification settings, these decision boundaries are already existing and predefined by the prediction classes. Thus, in classification settings, there is less need to consider multiple baselines to answer different questions. In fact, for classification tasks, a single baseline should suffice. The choice of this baseline is quite important, and as Sundararajan et al. (2017) suggested, it should be chosen so that it corresponds to a uniform distribution of baseline likelihoods for all classes, or in simple words, it contains no signal or information.  In this way, the attribution for any class will be a function only of the input, without the presence of artifacts originating from considering a baseline that is informative (Sundararajan et al., 2017).

## 5. Summary

In this study, we highlight our "lesson learned" that the attribution task, i.e., attributing a model's certain output to the corresponding input, does not have a single solution. Rather, the attributions and their interpretation depend greatly on the choice of the baseline choice; a highly overlooked issue in the literature. We illustrated this in a climate prediction task, where a fully connected network was trained to predict the ensemble- and global-mean temperature of annual temperature maps from a large ensemble of climate simulations. Our results make clear that when considering different baselines, attributions differ substantially both in magnitude and in spatial patterns.

We suggest that the dependence of the attribution task on the baseline choice is actually beneficial, since we can use different baselines to gain insights regarding different science questions, of varying complexity. We also highlight that in regression settings, the issue of the baseline needs to be cautiously considered to avoid misidentifying sources of predictability and/or artificially introducing inter-method discrepancies in XAI applications. In classification settings, a single, non-informative baseline should suffice, since the decision boundaries are predefined by the prediction classes. Nevertheless, in all XAI applications, the baseline needs to be carefully chosen and explicitly stated.

**Acknowledgments**

This work was supported in part by the National Science Foundation under Grant No. OAC-1934668. I. E. also acknowledges support by the National Science Foundation under AI Institute Grant No. ICER-2019758. The authors would also like to thank the efforts of the CESM2 Large Ensemble team for making their data publicly available and the IBM Center for Climate Physics in South Korea for providing supercomputing resources.

**Data availability**

The CESM2-LE model output is publicly available through https://www.cesm.ucar.edu/projects/community-projects/LENS2/data-sets.html . The code to reproduce the presented results will become publicly available upon acceptance of the manuscript and before publication.